\documentclass[aps, prd, reprint, amsfonts, amssymb, amsmath, showkeys, nofootinbib, twoside, superscriptaddress, longbibliography]{revtex4-1}
\usepackage{caption, subcaption}
\usepackage[pdftex, pdftitle={Article}, pdfauthor={Author}]{hyperref} 
\usepackage{amsthm}
\usepackage{mathtools}
\usepackage{acronym}
\usepackage{xcolor}
\usepackage{graphicx}
\usepackage{xspace}
\usepackage[left=23mm,right=13mm,top=35mm,columnsep=15pt]{geometry} 

\newcommand{\pastro}{\ensuremath{p_\mathrm{astro}}\xspace{}}
\newcommand{\pycbc}{\textsc{PyCBC}\xspace{}}
\newcommand{\gstlal}{\textsc{GstLAL}\xspace{}}
\newcommand{\lambs}{\Lambda_\mathrm{s}\xspace{}}
\newcommand{\lambn}{\Lambda_\mathrm{n}\xspace{}}

\newcommand{\toypipe}[1]{\textit{Pipeline #1}}
\newcommand{\boldtext}[1]{#1}

\begin{document}
\title{A Unified \pastro{} for Gravitational Waves: Consistently Combining Information from Multiple Search Pipelines}

\author{Sharan Banagiri}
    \email[Email: ]{sharan.banagiri@northwestern.edu}
    \affiliation{Center for Interdisciplinary Exploration and Research in Astrophysics, Northwestern University, 1800 Sherman Avenue, Evanston, Illinois 60201, USA } 

\author{Christopher~P.~L.~Berry}
    \affiliation{SUPennsylvania, School of Physics and Astronomy, University of Glasgow, Glasgow G12 8QQ, United Kingdom} 

\author{Gareth~S.~Cabourn~Davies}
\affiliation{University of Portsmouth, Portsmouth PO1 3FX, United Kingdom} 

\author{Leo Tsukada}
\affiliation{Department of Physics, The Pennsylvania State University, University Park, Pennsylvania 16802, USA}
\affiliation{Institute for Gravitation and the Cosmos, The Pennsylvania State University, University Park, Pennsylvania 16802, USA} 

\author{Zoheyr Doctor}
\affiliation{Center for Interdisciplinary Exploration and Research in Astrophysics, Northwestern University, 1800 Sherman Avenue, Evanston, Illinois 60201, USA } 

\date{\today} 

\begin{abstract}
Recent gravitational-wave transient catalogs have used \pastro{}, the probability that a gravitational-wave candidate is astrophysical, to select interesting candidates for further analysis. 
Unlike false alarm rates, which exclusively capture the statistics of the instrumental noise triggers, \pastro{} incorporates the rate at which triggers are generated by both astrophysical signals and instrumental noise in estimating the probability that a candidate is astrophysical. 
Multiple search pipelines can independently calculate \pastro{}, each employing a specific data reduction. 
While the range of \pastro{} results can help indicate the range of uncertainties in its calculation, it complicates interpretation and subsequent analyses. We develop a statistical formalism to calculate a \emph{unified} \pastro{} for gravitational-wave candidates, consistently accounting for triggers from all pipelines, thereby incorporating extra information about a signal that is not available with any one single pipeline. 
We demonstrate the properties of this method using a toy model and by application to the publicly available list of gravitational-wave candidates from the first half of the third LIGO-Virgo-KAGRA observing run. 
Adopting a unified \pastro{} for future catalogs would provide a simple and easy-to-interpret selection criterion that incorporates a more complete understanding of the strengths of the different search pipelines

\end{abstract}


\maketitle

\acrodef{gw}[GW]{gravitational wave}
\acrodef{cbc}[CBC]{compact binary coalescence}
\acrodef{ligo}[LIGO]{Laser Interferometer Gravitational-Wave Observatory}
\acrodef{far}[FAR]{false alarm rate}
\acrodef{lvk}[LVK]{LIGO-Virgo-KAGRA}
\acrodef{bns}[BNS]{binary neutron star}
\acrodef{bbh}[BBH]{binary black hole}
\acrodef{nsbh}[NSBH]{neutron star--black hole binary}
\acrodef{o3}[O3]{third observing run}
\acrodef{kde}[KDE]{kernel density estimation}

\defcitealias{fgmc:2015}{FGMC}

\section{Introduction} 
\label{sec:intro}

The detection of \acp{gw}~\cite{LIGOScientific:2016aoc,LIGOScientific:2021djp} by the \ac{ligo}~\cite{LIGOScientific:2014pky} and Virgo~\cite{VIRGO:2014yos} detectors is the culmination of decades of research. Not only are sensitive detectors needed to the measure the \ac{gw} signals~\cite{Pitkin:2011yk,KAGRA:2013rdx}, but sophisticated data analysis is needed to distinguish astrophysical signals from detector noise~\cite{LIGOScientific:2019hgc}. 
In the case of transient signals, such as those from \acp{cbc}, detection algorithms identify candidate signals by matching template signals to the data~\cite{LIGOScientific:2016vbw} or by looking for coherent signals in multiple detectors~\cite{LIGOScientific:2016fbo}. 
Only after identifying \ac{gw} candidates in detector data can we start to \boldtext{understand} the astrophysical population of \ac{gw} sources.

A fundamental question in any data analysis problem is the veracity of the signal or the effect it is considering. 
The statistical significance of a transient \ac{gw} is generally assessed by calculating how likely the data would appear due to noise fluctuations. 
This may be quantified by the \ac{far} or by a p value. Statistics like \ac{far} are particularly well suited for making a first detection, where we do not know the population of signals. 
However, a more complete assessment of the probability that a candidate is real can be obtained by considering not only the \ac{far} but also the \emph{true} alarm rate (i.e., how often the algorithm would find identify a real signal) whose calculation requires knowledge of the source population and our sensitivity to their signals, and hence is subject to additional uncertainties not inherent in the calculation of the \ac{far} or p value. 
By combining the false and true alarm rates, we can calculate the probability of astrophysical origin \pastro{} for a candidate~\cite{fgmc:2015,Guglielmetti:2009wm}. 
In the GW literature, \pastro{} was first used to estimate the astrophysical probability of GW150914 and the then-considered marginal trigger LVT151012~\cite{LIGOScientific:2016kwr,LIGOScientific:2016ebi} (now known as GW151012~\cite{LIGOScientific:2018mvr}). 
The probability of astrophysical origin \pastro{} directly addresses the key question of how probable a candidate is to be real accounting for both our understanding of detector sensitivity and the population of \ac{gw} sources.

We are now in the era of having many \ac{gw} candidates, such that we are building an understanding of the source population~\cite{LIGOScientific:2021psn}. 
\acp{gw} catalogs from both the \ac{lvk} Collaboration~\cite{LIGOScientific:2018mvr,LIGOScientific:2021djp} and independent teams~\cite{Nitz:2021zwj,Olsen:2022pin} now commonly use \pastro{} as a criterion to identify interesting candidates for further analysis.
By the end of their \ac{o3}, the \ac{lvk} Collaboration have reported {$90$}  \ac{cbc} candidates \boldtext{ with $\pastro{} > 0.5$}~\cite{LIGOScientific:2018mvr,LIGOScientific:2020ibl,LIGOScientific:2021usb,LIGOScientific:2021djp}. 
Despite the additional uncertainties inherent in its calculation, \pastro{} is important to consider now that we now have observations from multiple observing runs of different sensitivity~\cite{KAGRA:2013rdx} and observations of different types of sources~\cite{LIGOScientific:2016aoc,LIGOScientific:2017vwq,LIGOScientific:2021qlt,LIGOScientific:2021djp}. 
At a given \ac{far}, a candidate from the most recent \ac{o3} run~\cite{aLIGO:2020wna,Virgo:2022ypn} is more probable to be real than a candidate from the (less sensitive) first observing run~\cite{LIGOScientific:2016emj}. 
Similarly, at a given \ac{far}, a candidate consistent with originating from a $30 M_{\odot}+30 M_{\odot}$  \ac{bbh} merger is more probable to be real than a candidate consistent with a \ac{bns} origin~\cite{Kapadia:2019uut} because the former can be detected from a greater distance and so are more frequently observed~\cite{LIGOScientific:2021djp}.  
Since \pastro{} accounts for the true alarm rate, it can naturally be used when compiling a heterogeneous catalog of diverse sources from different observing runs.

During \ac{o3}, the \ac{lvk} Collaboration used four search pipelines for its final analysis~\cite{LIGOScientific:2021djp}, with each calculating its own \pastro{}: there are three \ac{cbc} matched-filtered pipelines (\gstlal~\cite{Messick:2016aqy, Sachdev:2019vvd, Hanna:2019ezx, Cannon:2021, Kapadia:2019uut}, \pycbc~\cite{Allen:2005fk, DalCanton:2014hxh, Usman:2015kfa, Nitz:2017svb, Davies:2020tsx, dent_pastro} and \textsc{MBTA}~\cite{Adams:2015ulm, Aubin:2020goo,Andres:2021vew}), and non-template-based search pipeline (\textsc{cWB}~\cite{klimenko:2004, Klimenko:2011hz, Klimenko:2015ypf}) that makes minimal assumptions about the transient signals. 
Multiple analyses with different choices help to explore the \ac{gw}-signal space more thoroughly, making it easier to find a diverse range of signals. 
Unfortunately, this complicates the use of \pastro{} when interpreting current \ac{gw} candidates in three ways. 
First, there are differences in the assumptions about the source population between analyses that mean that results are not directly comparable. 
Second, as illustrated in Fig.~6 of GWTC-3~\cite{LIGOScientific:2021djp}, pipelines can have different relative sensitivities, which may provide valuable insights into whether the candidate is real. 
This information is not incorporated in individual \pastro{} estimates. 
Third, it means there is no single assessment of the significance of a candidate, complicating the calculation of contamination and assessment of search sensitivity (and hence selection effects, which are essential for population studies). 
Indeed, as can be seen in Table~1 of GWTC-3~\cite{LIGOScientific:2021djp}, pipelines can often find a different set of triggers over the same stretch of data and can sometimes calculate significantly different \pastro{} values for the same candidate. 
To fully make use of the different search pipelines, it is necessary to combine their results to produce a single \pastro{} value for each candidate.

\boldtext{Some early work in combining results from multiple pipelines was done in the pre-detection era. 
For example, developing a way to estimate generalized frequentist upper limits with multiple pipelines~\cite{Sutton:2009up}, and developing unified detection statistics~\cite{Biswas:2012ty}. 
In this paper,} we present a framework that can combine the results from multiple detection algorithms to produce a unified \pastro{} employing information about the correlation between pipelines. 
This means that catalogs can be produced consistently using a simple threshold on this statistic and that downstream users may more straightforwardly calculate the contamination fraction for such a catalog. 
This statistic is amenable to use in population inferences that incorporate low-significance candidates~\cite{Galaudage:2019jdx, Roulet:2020wyq}. 
Crucially, as the formalism accounts for the correlations between different pipeline outputs (or lack thereof), we can make use of the full ensemble of results when evaluating whether a candidate is real.
We present an illustration of the properties of our \pastro{} formalism, and provide a proof-of-concept example of its use with real \ac{gw} data from the first half of \ac{o3} (henceforth called O3a) analyzed for GWTC-2.1~\cite{LIGOScientific:2021usb}.

In Sec.~\ref{sec:pastro} we define \pastro{} from first principles and describe its connection with the Poisson mixture-model formalism, first developed by Farr \textit{et al}.~\cite{fgmc:2015} (henceforth called the FGMC method) in the \ac{gw} context. 
Section~\ref{sec:joint-pastro} extends this formalism to calculate unified \pastro{} from multiple search pipelines. We demonstrate some of its useful properties by the means of a simple toy model in Sec.~\ref{sec:toy-model}, followed by an illustrative application to triggers from GWTC-2.1 in Sec.~\ref{sec:realdata}. 
This illustration highlights the tools that need to be developed in order to produce a reliable unified \pastro{} for use in future \ac{gw} catalogs. 
We discuss the applications and extensions of our formalism in Sec.~\ref{Sec:Discussion}.

\section{Defining \pastro} 
\label{sec:pastro}

In this section, we present a pedagogical derivation for \pastro{} as commonly seen in literature~\cite{fgmc:2015, Kapadia:2019uut, LIGOScientific:2016kwr,LIGOScientific:2016ebi}, connecting it the Poisson mixture-model FGMC formalism. 
In Sec.~\ref{sec:joint-pastro}, we explain how to extend this for the case of multiple search pipelines.

The primary goal of a search pipeline is to identify candidates of astrophysical origin. 
Pipelines usually accomplish this by maximizing some statistics over a small stretch of data, and identifying a particular candidate, which we will call a trigger. 
We assume that the stretch of data contains at most one signal. 
Every trigger has a detection statistic $x$ associated with it by the search pipeline.

The distribution of the detection statistic under pure noise lets us calculate the \ac{far} distribution of the pipeline. In the context of \ac{gw} analysis, this can be done by the bootstrapping method of time slides, where the data stream from one detector is shifted with respect to another by more than the light-travel time between them~\cite{Babak:2012zx}, or by constructing a model of the noise using triggers that are not coincident between detectors~\cite{Cannon:2012zt}. 

We define \pastro{} as the probability that a particular trigger is caused by an astrophysical signal, as opposed to a noise fluctuation or terrestrial contamination. 
Hence \pastro{} is properly associated with the trigger and its detection statistic (one specific reduction of the data) rather than directly the underlying data itself. 
When different pipelines analyze the data, they make different analysis choices which give them differing sensitivities, and so it is possible that they yield different triggers and \pastro{} values. 

Since a trigger can be caused by an astrophysical signal or by noise, \pastro{} depends on the posterior probability for these two hypotheses. 
We define $p (\mathcal{S} | x, \Phi_{\mathrm{s}}, \Phi_{\mathrm{n}})$ as the posterior probability for the signal hypothesis $\mathcal{S}$, and $p (\varnothing | x, \Phi_{\mathrm{n}})$  as the posterior probability for the noise hypothesis $\varnothing$. 
Both these probabilities are conditioned on the detection statistic of the trigger $x$, and assume some signal and noise parameters, $\Phi_{\mathrm{s}}$ and $\Phi_{\mathrm{n}}$. 
We can then write \pastro{} for the trigger as the normalized probability that it is astrophysical, 
\begin{equation}
    \pastro(x) = \frac{p (\mathcal{S} | x, \Phi_{\mathrm{s}}, \Phi_{\mathrm{n}})}{p (\mathcal{S} | x, \Phi_{\mathrm{s}}, \Phi_{\mathrm{n}}) + p (\varnothing | x, \Phi_{\mathrm{n}}) }. 
\label{Eq:pastro_def}
\end{equation} 
We shall suppress the signal and noise parameters from here on for the purpose of clarity. 
A value $\pastro{} = 1$ implies perfect confidence about the presence of a signal in an underlying segment of data, while $\pastro{} = 0$ implies perfect confidence of its absence, and $\pastro{} = 0.5$ implies no preference between the signal and noise hypothesis. 

The posterior probabilities for two cases are difficult to compute directly. 
What is more easily accessible is the likelihood of the trigger $x$, under the signal and noise hypotheses. 
Therefore using Bayes' theorem, we write $p (\mathcal{S} | x)$ and $p (\varnothing | x)$ as 
\begin{equation}
\begin{split}
        p (\mathcal{S} | x)  = \frac{\pi_{\mathrm{s}}  p (x | \mathcal{S} )}{Z(d)}, \\
        p (\varnothing | x) = \frac{\pi_{\mathrm{n}}  p (x | \varnothing )}{Z(d)}, \\
\end{split}
\label{Eq:BayesTheorem}
\end{equation}
where $p (x | \mathcal{S} )$ and $p (x | \varnothing )$ are the likelihoods for getting a trigger with statistic $x$ under the signal and noise hypothesis, respectively; $\pi_{\mathrm{s}}$ and $\pi_{\mathrm{n}}$ are the corresponding priors, and $Z(d)$ is the Bayesian evidence. 
We can then write \pastro{} as 
\begin{equation}
    p_{\rm astro} (x) = \frac{\pi_{\mathrm{s}} p(x | \mathcal{S})}{\pi_{\mathrm{s}}  p (x | \mathcal{S}) +  \pi_{\mathrm{n}}   p (x | \varnothing ) }.
    \label{Eq:pastro_unmarg}
\end{equation}

Suppose that we analyze a long stretch of data with a single search pipeline. Let $R_\mathrm{s}$ and $R_\mathrm{n}$ be the rates of astrophysical and noise-only triggers, respectively, assuming a given detector sensitivity, operating characteristics, and algorithmic choice (including any preliminary thresholds that are applied) made by the search pipeline. 
We define the signal and noise count parameters, $\lambs{}$ and $\lambn{}$, respectively. 
These are the mean number of astrophysical and noise triggers (above the predefined thresholds) for an observing duration $T$ (not necessarily the number of triggers in any particular realization of the data) such that 
\begin{equation}
    \lambs{} = R_\mathrm{s} T \qquad \text{ and } \qquad \lambn{} = R_\mathrm{n} T. 
\end{equation}
The total number of triggers ${N}$ will follow a Poisson distribution with a count parameter $\Lambda = \lambs{} + \lambn{} $:
\begin{equation}
    p({N} | \lambs{} , \lambn{} ) = \frac{ \left(\lambs{} + \lambn{} \right)^{N} \, e^{-\left(\lambs{} + \lambn{} \right)} }{{N}!}.
    \label{Eq:Poisson_stats}
\end{equation}

For computational reasons, it is common to require that the trigger has to pass some primary statistical threshold, before being used for further analysis. 
For example, the GWTC-2.1~\cite{LIGOScientific:2021usb} and GWTC-3~\cite{LIGOScientific:2021djp} analyses considered only triggers that had a \ac{far} $\leq 2~\mathrm{day}^{-1}$. 
Our formalism does not make any strong assumptions about the preliminary threshold, as long as it is consistently used among all pipelines. Thresholds are incorporated by suitably defining the counts $\lambs{}$ and $\lambs{}$ as for triggers that pass the threshold.

The count parameters are generally assumed to be unknown and have to be measured empirically using the triggers. 
To do this we use the Poisson mixture-model FGMC formalism:
\begin{equation}
     p(\{x\} | \lambs{}, \lambn{}, N ) = \prod_i^N \left [  \pi_{\mathrm{s}}  p({x_i} | \mathcal{S} ) + \pi_{\mathrm{n}} p({x_i} | \varnothing) \right], 
     \label{Eq:mix_model}
\end{equation}
where $\{x\}$ denotes the set of all triggers found in the data. 
Conditioned upon $\lambs{}$ and $\lambn{}$, the prior probabilities are just the relative rate of the triggers of that category, 
\begin{equation}
    \pi_{\mathrm{s}} = \frac{\lambs{}}{\lambs{} + \lambn{}}, \qquad  \pi_{\mathrm{n}} = \frac{\lambn{}}{\lambs{} + \lambn{}}. 
    \label{Eq:Prior_counts}
\end{equation}
Therefore, Eq.~\eqref{Eq:mix_model} becomes
\begin{equation}
     p(\{x\} | \lambs{}, \lambn{}, N ) = \frac{\prod_i^N \left[  \lambs{}  p({x_i} | \mathcal{S} ) +\lambn{} p({x_i} | \varnothing) \right] } {(\lambs{} + \lambn{})^{{N}}}.
\end{equation}

Using Bayes' theorem, we can relate the posterior distribution on the count parameters with the the Poisson likelihood, 
\begin{equation}
\begin{split}
        p(\lambs{}, \lambn{} | \{x\},  {N}) & \propto p( \{x\},  {N} | \lambs{}, \lambn{}) \, \pi (\lambs{}, \lambn{}) \\
        & \propto  p( \{x\} |  {N} , \lambs{}, \lambn{}) \, p( {N} | \lambs{}, \lambn{})  \pi(\lambs{}, \lambn{}),
\end{split}
\end{equation}
where $\pi (\lambs{}, \lambn{})$ is the prior on the mean count. 
Putting this together with Eq.~\eqref{Eq:Poisson_stats}, we obtain the posterior distributions for $\lambs{}$ and $\lambn{}$, 
\begin{equation}
\begin{split}
        p(\lambs{}, \lambn{} |\{x\} , {N}) \propto \, & e^{-(\lambn{} + \lambs{})} \pi(\lambs{}, \lambn{}) \times \\ & \prod_i^N \left \{  \lambs{}  p({x_i} | \mathcal{S} ) +\lambn{} p({x_i} | \varnothing) \right \}.
\end{split}
\label{Eq:FGMC_estimate}
\end{equation}

Finally, combining Eq.~\eqref{Eq:pastro_unmarg} and Eq.~\eqref{Eq:Prior_counts}, we can calculate a \pastro{} that marginalizes over the posterior of $\lambs{}$ and $\lambn{}$, 
\begin{equation}
    p_{\rm astro} (x) = \int \mathrm{d} \lambs{} \mathrm{d}\lambn{} \frac{\lambs{} p(x | \mathcal{S}) \, p(\lambs{}, \lambn{}| \{x\}, N) }{\lambs{}  p (x | \mathcal{S}) +  \lambn{}   p (x | \varnothing ) }.
    \label{Eq:pastro_marg}
\end{equation}
This fundamental framework has been used by all search pipelines both by the \ac{lvk} Collaboration and other groups in compiling \ac{gw} catalogs~\cite{LIGOScientific:2016kwr, LIGOScientific:2016ebi, LIGOScientific:2021djp, LIGOScientific:2021usb, Olsen:2022pin, Venumadhav:2019lyq, Nitz:2020oeq, Nitz:2021uxj, Nitz:2021zwj}.

\section{Towards a unified \pastro}
\label{sec:joint-pastro}

We now develop a way to combine information from multiple pipelines for calculating a unified \pastro. 
Suppose that we have several search pipelines that yield triggers after running over the same underlying data. 
Consider any two corresponding triggers $x^\alpha$ and $x^\beta$ from pipelines $\alpha$ and $\beta$; as these triggers are produced by different pipelines they may be associated with different statistics, but these statistics will be correlated depending upon the relative sensitivities of the two pipelines. 
Understanding the correlations between pipelines is key to understanding how to construct a unified \pastro{}.

The correlations between different search pipelines are not typically accounted for when compiling \ac{gw} results. 
In some analyses, the \ac{far} is multiplied by a trials factor equal to the number of different searches~\cite{LIGOScientific:2017zid,LIGOScientific:2021tfm}. 
However, this is a conservative choice: it would be correct if the results were noise triggers that were uncorrelated, but in general, we would expect some correlation since search pipelines are searching for similar signals in the data. 
This correlation should reduce the effective trials factor (in the limit of running two identical pipelines there would be no need to add a trials factor). 
Our framework accounts for correlations in both noise triggers and signal triggers to construct a unified \pastro{}.

Let us define $\vec{x} = \{ x^\alpha, x^{\beta}, .... \} $ as the triggers that correspond with each other from a series of different pipelines. 
For the rest of this paper, a vector usually means a vector of pipeline outputs, such that $\{\vec{x}\}$ is the set of all triggers from all the pipelines in the stretch of data being analyzed, and a latin index is used to indicate the individual data segment or trigger.
As in Eq.~\eqref{Eq:BayesTheorem}, the probability $p (\mathcal{S} | \vec{x})$ is given by, 
\begin{equation}
    p (\mathcal{S} | \vec{x}) \propto \lambs{}   p (\vec{x} | \mathcal{S} ). 
\end{equation}
Similarly, for the noise hypothesis, 
\begin{equation}
        p (\varnothing | \vec{x}) \propto \lambn{}   p (\vec{x} | \varnothing ). 
\end{equation}
Here, $p (\vec{x} | \mathcal{S} )$ and $p (\vec{x} | \varnothing )$ are joint likelihoods for obtaining $\vec{x} = \{ x^\alpha, x^{\beta}, .... \}$ and are dependent on the aforementioned correlations between trigger statistics across pipelines. 
If we learn these joint likelihood distributions, using simulated signals and noise triggers, we can calculate a unified \pastro. 

While several methods to learn arbitrary distributions exist, we will primarily use \ac{kde} methods from \textsc{scikit-learn}~\cite{scikit-learn:2011} to learn $p (\vec{x} | \mathcal{S} )$ and $p (\vec{x} | \varnothing )$. The choice of method used to reconstruct the distribution is important, and the distributions must be properly characterized to successfully calculate a reliable \pastro{}. 
However, for the purposes of illustrating the framework needed to compute a unified \pastro, this method may be treated as a black box.

Once the distribution of these correlations is learned, one can modify Eq.~\eqref{Eq:FGMC_estimate} for a joint FGMC estimate 
\begin{equation}
\begin{split}
        p(\lambs{}, \lambn{} |\{\vec{x}\} , {N}) \propto \, & e^{-(\lambn{} + \lambs{})} \pi(\lambs{}, \lambn{}) \times \\ & \prod_i^N \left \{  \lambs{}  p({\vec{x}_i} | \mathcal{S} ) +\lambn{} p({\vec{x}_i} | \varnothing) \right \}.
\end{split}
\label{Eq:joint_FGMC}
\end{equation} 
Equation~\eqref{Eq:pastro_marg} can then be modified to calculate a unified \pastro{} marginalized over signal and noise counts, 
\begin{equation}
    p_{\rm astro} (\vec{x}) = \int \mathrm{d} \lambs{} \, \mathrm{d}\lambn{}  \frac{  \lambs{} p(\vec{x} | \mathcal{S}) \, p(\lambs{}, \lambn{} |\{\vec{x}\} , {N})}{\lambs{}  p (\vec{x} | \mathcal{S}) +  \lambn{}   p (\vec{x} | \varnothing ) }.
    \label{Eq:joint_pastro_marg}
\end{equation}
This equation therefore incorporates correlations between pipelines through the joint likelihoods $p (\vec{x} | \mathcal{S})$ and $p (\vec{x} | \varnothing )$.

The role of correlations in calculating \pastro{} may be understood by considering a few idealized cases. 
If we had two pipelines that looked for the same type of signal, but had different sensitivities to noise triggers, we might expect to be more certain in a candidate (assign a higher \pastro{}) if there are corresponding triggers from both pipelines. 
However, if a candidate was identified by one pipeline, and not by another which has greater sensitivity to that type of signal, we might expect to suspect it as a false alarm (assign a lower \pastro{}).
In the following subsections, we will demonstrate the properties of our unified \pastro{} in some illustrative cases.

\section{Toy Model} 
\label{sec:toy-model}

{In this section we illustrate and test the unified \pastro{} method using a simple toy model. 
In Sec.~\ref{sec:realdata}, we will then apply the method on real \ac{gw} data, albeit in a simplified analysis.} 

\begin{figure}
    \includegraphics[width=0.49\textwidth]{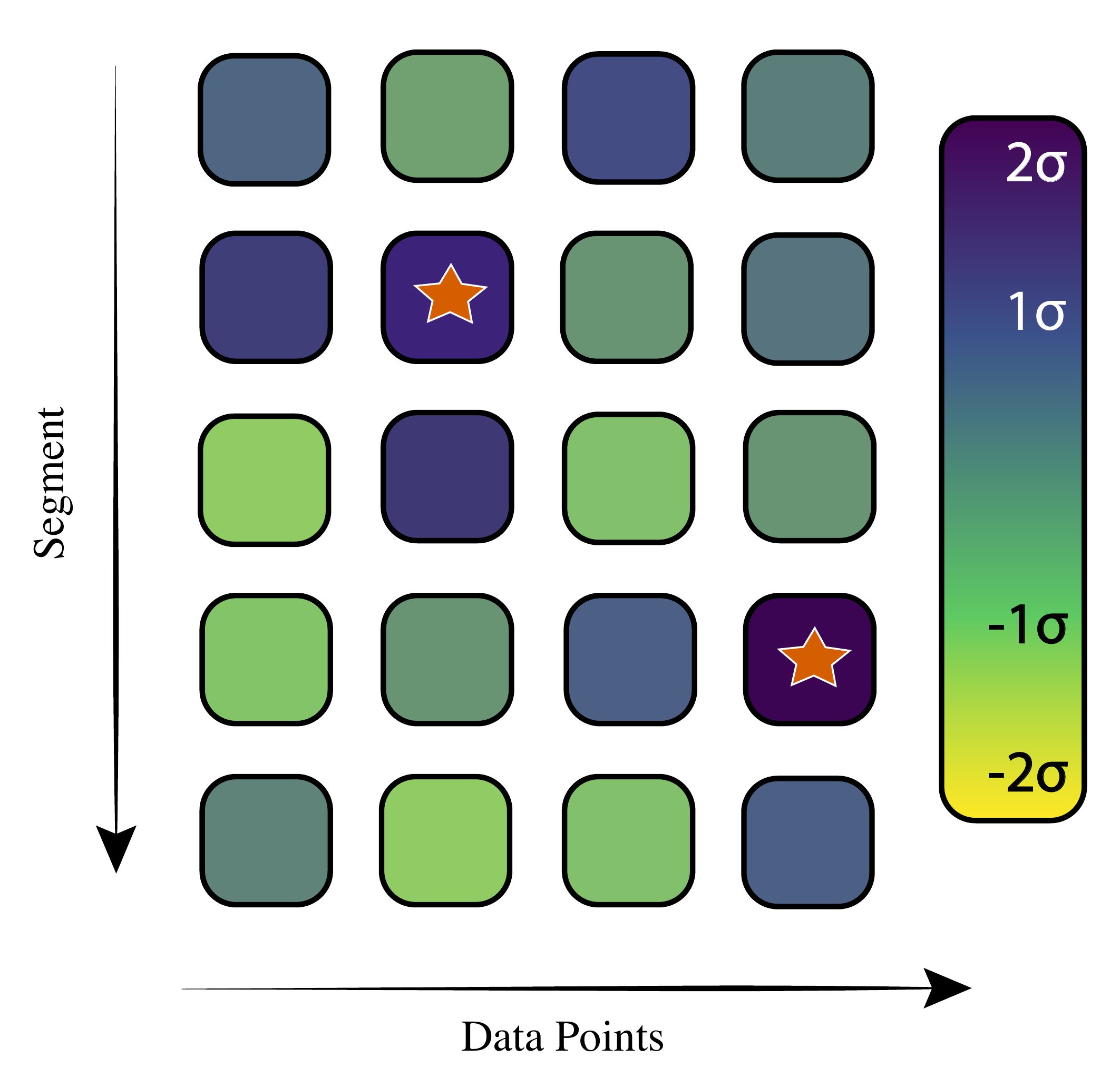}
    \caption{A schematic of the toy model data. 
    Each row is a segment, consisting of four data points indicated by the rounded rectangles. 
    The noise in the segments is drawn from a standard normal distribution, with its value indicated by the color scale of the rounded rectangle. 
    Segments 2 and 4 from the top in this example contain a signal, added randomly to one of the four data points indicated by the vermilion star. }

    \label{fig:toy_model_schematic}
\end{figure}

The toy model data is generated in the form of segments, each of which consists of four data points with values drawn from a standard normal distribution.
A segment might also contain a signal with a probability proportional to $\lambs{}$. 
Signals are simulated by adding a value $\lambda_\mathrm{s}$ to one of the data points in the segment. 
We assume that $\lambda_\mathrm{s}$ is known. 
The goal of an FGMC-type analysis is to estimate the fraction $\lambs{}$. 
Figure~\ref{fig:toy_model_schematic} demonstrates a schematic of the data generated in the toy model.

We construct four simple pipelines to analyze this data on a segment-by-segment basis. 
The pipelines are all based on a maximum-likelihood estimate assuming Gaussian noise statistics, and the pipelines all assume that the noise is drawn from a standard normal with $\sigma=1$. 
For each segment, the pipelines calculate the noise likelihood
\begin{equation}
    \mathcal{L}_\mathrm{n} \propto \exp\left[ - \frac{1}{2}\frac{\left(\sum_{i=1}^k x_i  \right)^2}{k \sigma^2} \right],
    \label{Eq:toymodel_noi_like}
\end{equation}
and the signal likelihood, assuming the signal value $\lambda_\mathrm{s}$ is known
\begin{equation} 
\mathcal{L}_\mathrm{s}
\propto \exp\left[ - \frac{1}{2}\frac{\left(\sum_{i=1}^k x_i - \lambda_{\mathrm{s}} \right)^2}{k \sigma^2} \right], 
    \label{Eq:toymodel_sig_like}
\end{equation}
where $x_i$ are the analyzed data points in the segment. 
The pipelines might use different number of data points, corresponding to the number $k \leq 4$. 
Each pipeline calculates a p value for every segment (under the noise hypothesis) which is the main input for the unified \pastro{} analysis.

The four pipelines are
\begin{enumerate}
    \item \toypipe{1}: This pipeline makes no assumption on which of the four data points in a segment has the signal, effectively marginalizing over the position of the signal. 
    Therefore, $k=4$ in Eq.~\eqref{Eq:toymodel_noi_like} and Eq.~\eqref{Eq:toymodel_sig_like}. 
    
    \item \toypipe{2}: This pipeline still assumes Gaussian statistics but only considers the data point with the loudest value of the four and calculates its likelihood. The is equivalent to setting $k= 1$ in Eq.~\eqref{Eq:toymodel_noi_like} and Eq.~\eqref{Eq:toymodel_sig_like}, but replacing $x_i$ with $\max\{x_i\}$. 
    This pipeline will be more effective at detecting a loud signal but will generally overestimate the number of signals.

    \item  \toypipe{3}: This only uses the first two data points of a segment. 
    Therefore, $k=2$ in Eq.~\eqref{Eq:toymodel_noi_like} and Eq.~\eqref{Eq:toymodel_sig_like}. 
    Since in our toy model, we know that each data point is equally probable to contain a signal, we expect that this pipeline will witness only about half of the total signals, and consequently its count parameter will be around half the true value. 
    Thereby, this is analogous to a search pipeline that will be sensitive to only a subset of \ac{gw} signals (e.g., those from high-mass systems). 
    
    \item  \toypipe{4}: Similar to \toypipe{3} except it only uses the last two data points. 
    Therefore \toypipe{4} is statistically independent from \toypipe{3} while still seeing only half of the total signals.  
    Combining this pipeline with \toypipe{3} is the simplest example of a correlation between pipelines.

\end{enumerate}
For each pipeline, we also estimate the single-pipeline signal and noise counts using the signal and noise likelihoods.

Using the p value from the pipelines as their detection statistics, we will calculate the joint likelihoods using a \ac{kde} as described in Sec.~\ref{sec:joint-pastro}. 
To do this we create a number of segments containing a signal to train a signal \ac{kde}, and similarly train a noise \ac{kde} using segments that do not contain a signal. 
Then for a simulation where $\Lambda_{\mathrm{s}}$ is unknown, we can score the outputs of the pipelines against the \ac{kde}s to perform a unified FGMC analysis using Eq.~\eqref{Eq:joint_FGMC}, and measure the noise and signal counts, $\lambn{}$ and $\lambs{}$. 

\begin{figure*}
    \centering
    \begin{subfigure}[b]{0.92\textwidth}            \includegraphics[width=0.9\textwidth]{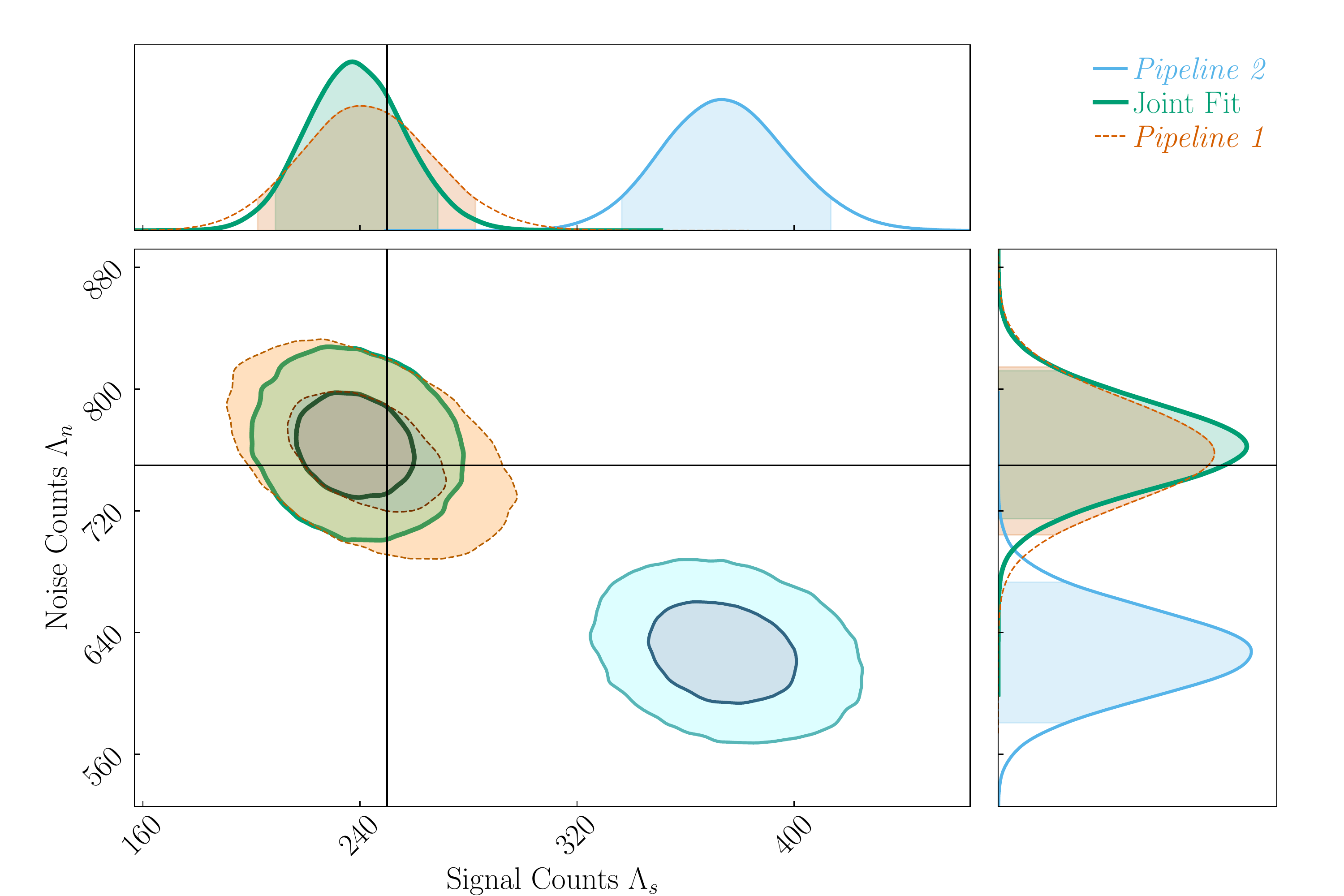}
    \caption{Analysis with \toypipe{1} and \toypipe{2}. 
    The horizontal and vertical axes are the measured signal and noise count parameters. 
    \toypipe{2} is clearly biased as it only uses the loudest data point of a segment for analysis, but the unified FGMC analysis \boldtext{performing a joint fit} can correct for it.} 
    \label{fig:pipe12}
    \end{subfigure}

    \begin{subfigure}[b]{0.92\textwidth}
    \includegraphics[width=0.9\textwidth]{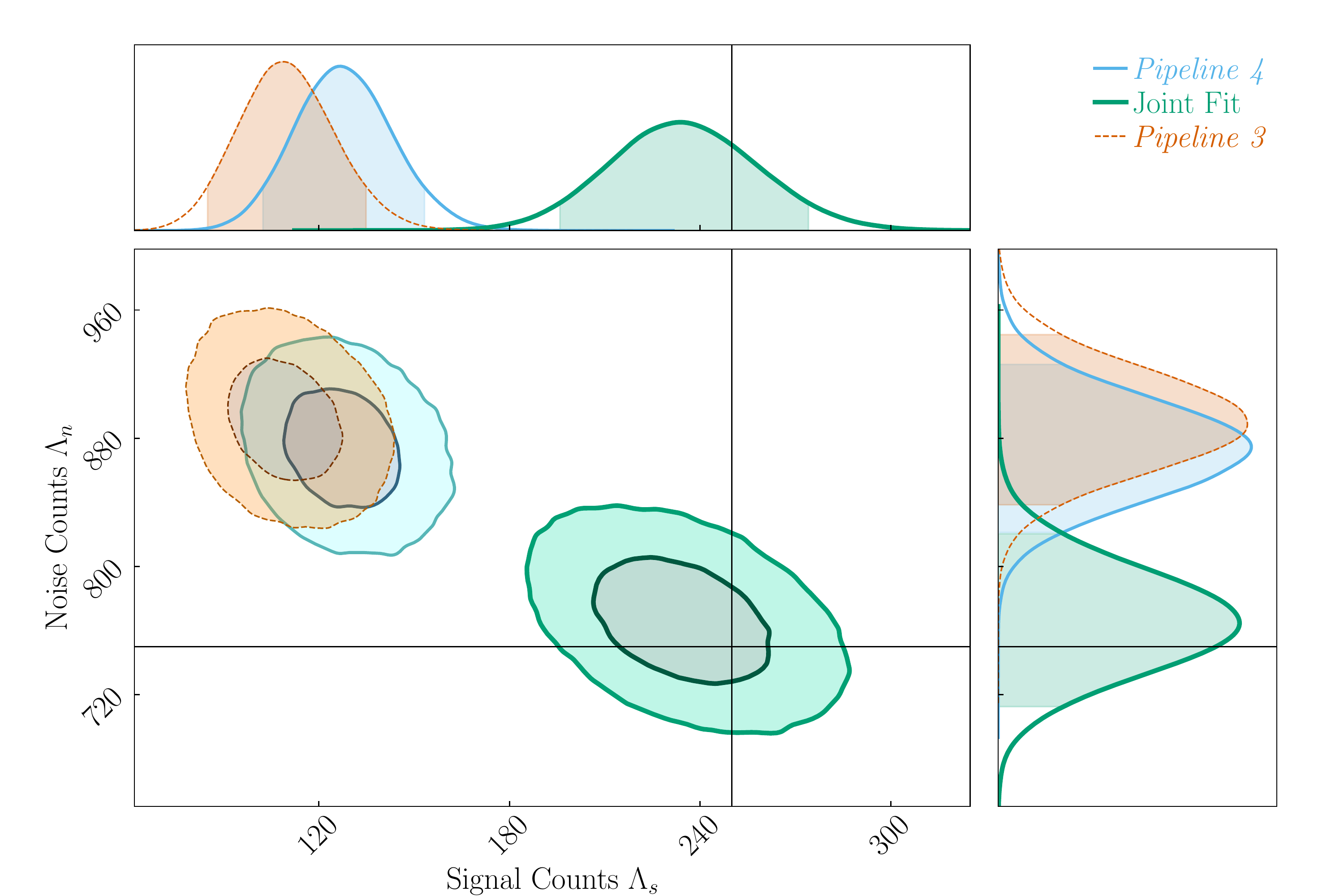}
    \caption{Analysis with \toypipe{3} and \toypipe{4}. While both pipelines only see half the simulated signals, the joint analysis learns to correct for this from the correlations in the joint signal distributions.} 
    \label{fig:pipe34}
    \end{subfigure}
    
    \caption{
    Toy-model exploration of the unified \pastro{} formalism. A signal value, $\lambda_\mathrm{s} = 3$ was used in both cases. In both figures, the \boldtext{solid vertical and horizontal lines} indicate the real number of segments with (and without) a signal. 
    \boldtext{The shaded regions in the marginal plots are $90\%$ uncertainty levels while the two-dimensional contours correspond to 50\% and 90\% levels.}}
    \label{fig:toy_model}
\end{figure*}

The algorithmic choices made by the pipelines mean that their results differ from each other, and can introduce a bias in their estimate of signal counts (and thereby \pastro{}). 
For instance, consider \toypipe{3}, which can only see about half of the total signals. 
When combining the output of the pipelines, the \acp{kde} do not know \textit{a priori} about these biases and will have to learn it from the triggers. In a joint analysis between \toypipe{3} and \toypipe{4},  the signal \ac{kde} can learn that only about half the analyzed segments will have a low p value in \toypipe{3} (i.e., when a signal is added to the first two data points) and that these are the segments in which \toypipe{4} will likely give a high p value (and vice versa). 

Figure~\ref{fig:toy_model} shows an FGMC analysis for two separate toy model analyses. 
In both Figs.~\ref{fig:pipe12} and~\ref{fig:pipe34}, we separately simulated $1000$ segments of data with about a quarter ($\Lambda_s=250$) of the segments having a signal with $\lambda_\mathrm{s}=3$. Figure~\ref{fig:pipe12} shows the signal and noise counts estimated by \toypipe{1} and \toypipe{2}, along with a unified fit. 
As expected, \toypipe{2} shows a bias in its recovery which nevertheless the joint analysis can correct for. 
Figure~\ref{fig:pipe34} shows the results of a toy-model analysis with \toypipe{3} and \toypipe{4}. 
Both pipelines only see about half of the simulated signals, by design. 
However, the joint analysis learns this bias through the signal \ac{kde} and can account for it. 

This section shows, using a simple example, how outputs from multiple pipelines can be combined using the unified formalism that we have developed. 
It also shows how certain kinds of biases can be corrected when multiple pipelines are joined together, by drawing information from the correlations (or lack thereof) between the outputs of the pipelines. 
These examples demonstrate the robustness of the unified \pastro{} method.

\begin{figure*}[hbtp]

    \centering
    \begin{subfigure}[b]{0.9\textwidth}
            \includegraphics[width=0.9\textwidth]{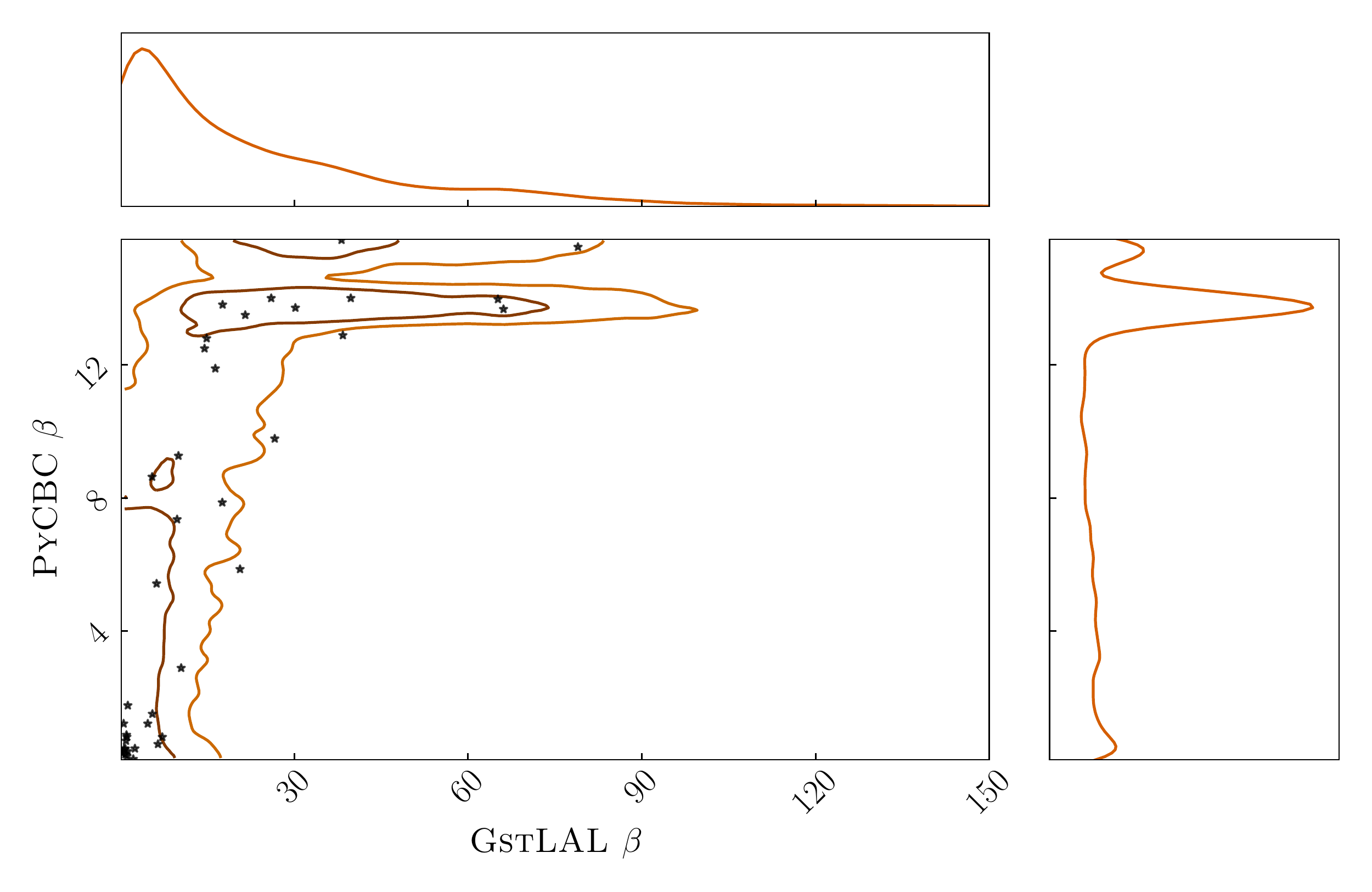}
            \caption{ The signal distribution $p(\vec{x} | \mathcal{S})$ learned using a \ac{kde} to the simulated signals that are commonly found by both \gstlal{} and \pycbc{}. 
            The stars are the O3a on-source triggers from GWTC-2.1 found by both pipelines~\cite{LIGOScientific:2021usb}. 
            The upper cutoff of the \pycbc{} distribution that is visible in the plot comes from the fact that the pipeline places a limit on the \ac{far} based on the number of time slides performed. 
            This translates to a cutoff at $\beta \approx 15.76$ \boldtext{and results in a bump in the \pycbc{} distribution at these values of $\beta$. }}
            \label{fig:joi_inj_kde}
    \end{subfigure}
    \centering
    \begin{subfigure}[b]{0.9\textwidth}
            \includegraphics[width=0.9\textwidth]{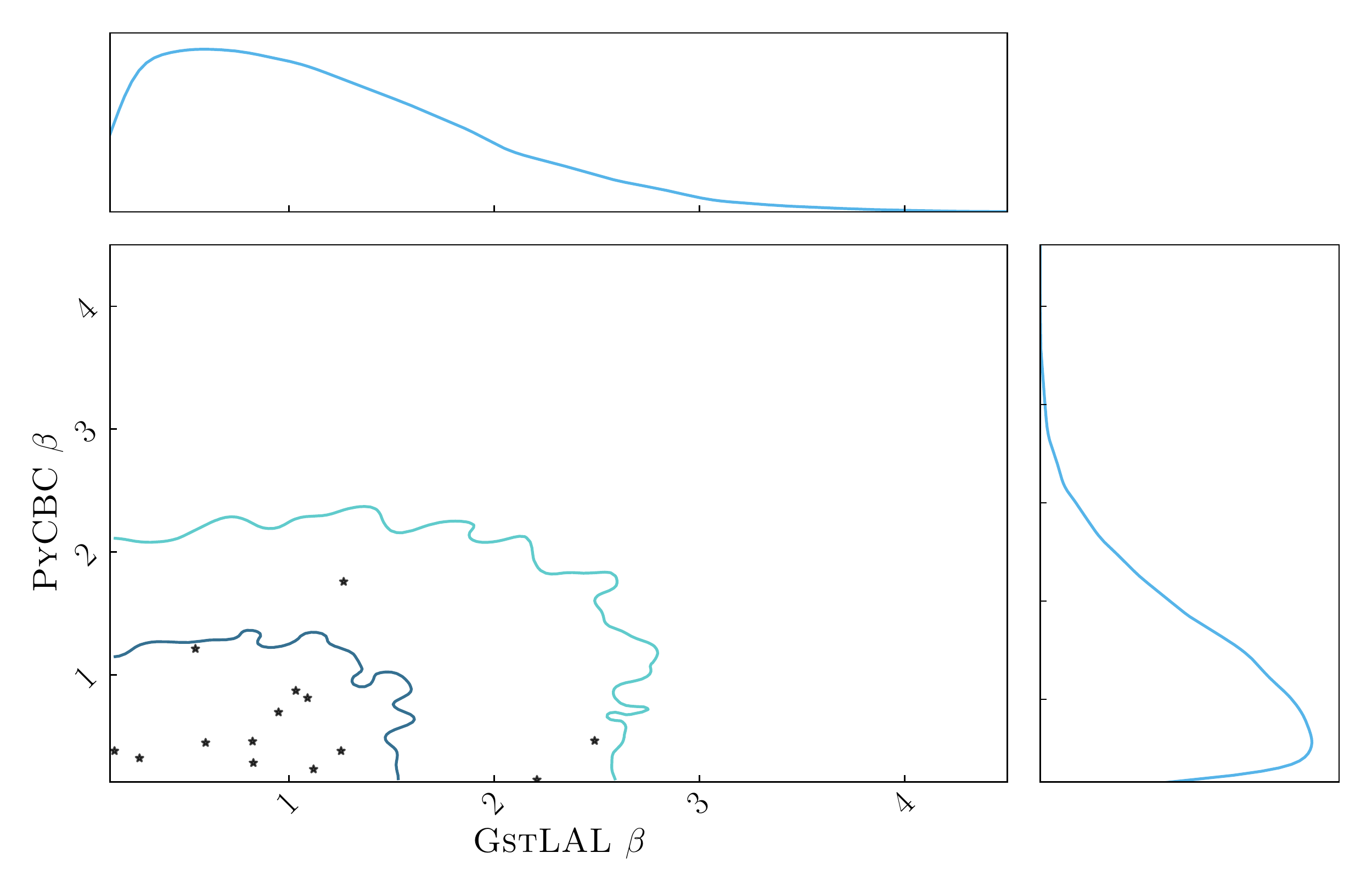}
            \caption{The noise distribution $p (\vec{x} | \varnothing )$ fit to a $2$-dimensional Gaussian to the common noise triggers. 
            The stars are the O3a on-source triggers from GWTC-2.1 found by both pipelines~\cite{LIGOScientific:2021usb}.}
            \label{fig:joi_noi_kde}
    \end{subfigure}
    \caption{Joint likelihood distributions for the signal and noise hypothesis. 
    The spans of the signal and noise distributions span are dissimilar. 
    The signal distribution extends to much larger $\beta$ values than the noise distribution. \boldtext{The contours correspond to $50 \%$ and $90 \%$ levels in both plots.}}
    \label{fig:joi_kde}
\end{figure*}

\begin{figure*}
    \centering
    \includegraphics[width=1.0\textwidth]{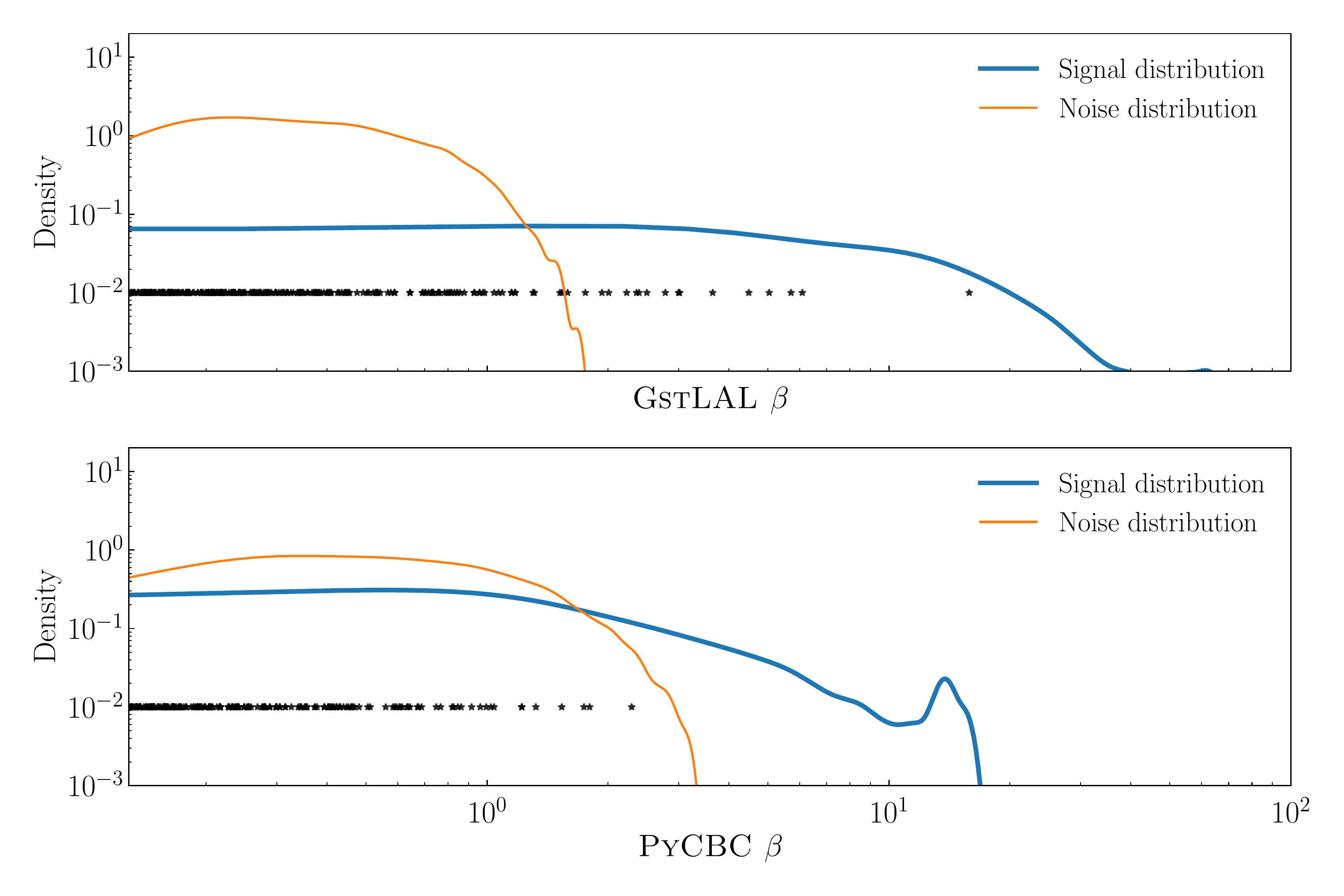}
    \caption{The distribution of triggers that were found by one pipeline only. 
    The top plot shows the noise and signal distribution of \gstlal{}, while the bottom plot shows the \pycbc{} distribution of triggers. 
    The black stars are again the on-source O3a triggers from GWTC-2.1 that have only been found by the corresponding pipeline~\cite{LIGOScientific:2021usb}.}
    \label{fig:marg_distribution}
\end{figure*}

\section{Application to GWTC-2.1} 
\label{sec:realdata}

We now test the unified \pastro{} method with real \ac{gw} triggers (henceforth \emph{on-source} triggers) by applying this to O3a results from GWTC-2.1~\cite{LIGOScientific:2021qlt}. 
GWTC-2.1 applies a preliminary cut of $\mathrm{FAR} \leq 2~\mathrm{day}^{-1}$, which we also adopt. While this analysis uses real data, it is only intended to be illustrative, and we adopt some simplifications for computational simplicity.

We use results from the \gstlal{}~\cite{Messick:2016aqy,Sachdev:2019vvd,Cannon:2021} and \pycbc{}~\cite{Usman:2015kfa,Nitz:2017svb,Davies:2020tsx} pipelines, and restrict ourselves to triggers that are at least Hanford-Livingston coincident (i.e., present in data from LIGO Hanford and Livingston at a minimum). 
\boldtext{GWTC-2.1~\cite{LIGOScientific:2021qlt} uses two versions of the \pycbc{} pipelines; referred to as \pycbc{}-\textsc{broad} and \pycbc{}-\textsc{BBH}; in this analysis we only use the former version, which we will simply refer to as \pycbc{}}.%
\footnote{\boldtext{Our approach could be used to combine the \pycbc{}-\textsc{broad} and \pycbc{}-\textsc{BBH} results too; we pick the \gstlal{} and \pycbc{}-\textsc{broad} results to illustrate our method to show how it can be used across different pipelines.}}
While our formalism is general, we focus solely on \ac{cbc} signals; a choice that is primarily driven by prudence, as we have only seen the population of \ac{cbc} signals so far~\cite{LIGOScientific:2021djp}. 
Furthermore, for the sake of simplicity we will only use \ac{bbh} signals for the training; this will also serve as a test to see if the method can distinguish sufficiently between signal types and to test overfitting. 
These simplifications make it easier to illustrate the behavior of the unified \pastro{}, and highlight the key considerations for an analysis to be performed for future \ac{gw} catalogs.

We use pipeline settings that are as similar as possible to the runs in GWTC-2.1~\cite{LIGOScientific:2021qlt}. 
A crucial factor to consider is that the search pipelines have to be run in a way that we can establish a one-to-one correspondence between their triggers so that we can learn the correlations between them.  
This is necessary even in the case where one pipeline registers a trigger and the other does not because the absence of the trigger is itself useful information. 
While one-to-one correspondence is straightforward for simulated signal triggers (described in Sec.~\ref{Sec:gwtc2.1_sig_kde}), it can be somewhat ambiguous to define in the case of noise triggers (implementation details are explained in Sec.~\ref{Sec:gwtc2.1_noise_kde}). 

In order to calculate a unified \pastro{}, Eq.~\eqref{Eq:joint_FGMC}, and Eq.~\eqref{Eq:joint_pastro_marg} require us to construct the joint distribution of a statistic from each pipeline. 
We choose to use the \ac{far} as the base statistic here as it is easily interpretable, commonly used by all pipelines and informative of the relative significance of triggers. 
However, the dynamic range of \acp{far} can be large, and we therefore define a statistic $\beta$ that is related to the logarithm of the inverse \ac{far},
\begin{equation}
    \beta(\mathrm{FAR}) = \log \left(1 + \frac{\kappa}{\mathrm{FAR}} \right),
\end{equation}
where $\kappa$ is a scaling constant, which we set to a value of $\kappa = 100~\mathrm{yr}^{-1}$. 
This number is set to give a good dynamic range to $\beta$ but \boldtext{our} results are insensitive to its actual value. 
Higher values of $\beta$ correspond to more significant candidates.

We construct three cases each for the noise and signal hypotheses, corresponding to triggers that are registered (i)  only by \gstlal, (ii) only by \pycbc, and (iii) by both pipelines. 
We find that separate modeling the three cases is necessary in order to achieve a faithful fit of signal triggers, which can cover a wide range on the $\beta$ space and show significant structure. In addition, this allows us to consider the case where one pipeline would not see a signal but the other does. For example, in the GWTC-2.1 analysis~\cite{LIGOScientific:2021usb}, \pycbc{} did not assign \acp{far} to events which triggered in a single detector (although this is now possible~\cite{Davies:2022thw}), but \gstlal{} did. 
In the future, a more sophisticated fitting method like a convolutional neural network or a random forest classifier might make multiple separate fits unnecessary. 

The distributions are usually normalized such that the total probability over its parameter space is one. 
Therefore, when considering multiple separate cases we \boldtext{need} to renormalize them to account for the relative probability of the class they represent. 
We do this by multiplying the output of the distribution with the fraction of signal or noise triggers that fall in that particular class. 
We find that getting a joint trigger is $\approx 104$ times more likely under the signal hypothesis than the noise hypothesis. 
Meanwhile, a \gstlal{}-only (\pycbc{}-only) trigger is $\approx 2.77$ ($\approx 3.87$) times more likely under the noise hypothesis than the signal hypothesis. 
The numbers quoted correspond to triggers that pass a $2~\mathrm{day}^{-1}$ \ac{far} threshold.

\begin{figure*}
    \centering
    \includegraphics[width=1.0\textwidth]{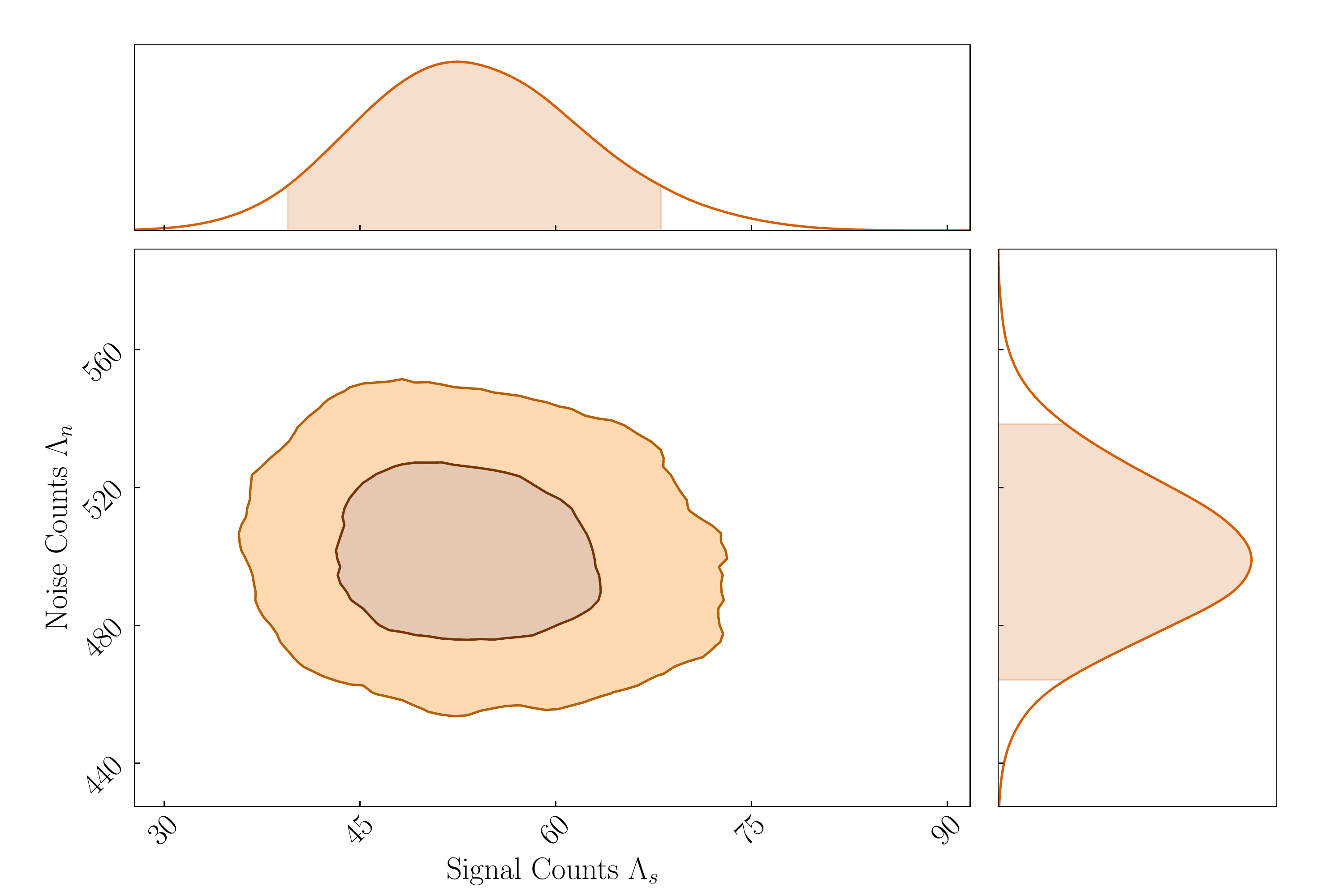}
    \caption{The posterior for the signal and noise counts in the joint analysis. 
    The shaded region in the one-dimensional posterior corresponds to $90\%$ uncertainty levels, while the contours in the two-dimensional posteriors are the $50\%$ and $90\%$ levels. 
    We recover a median value $\Lambda_s = 53^{+10}_{-13}$. }
    \label{fig:fgmc}
\end{figure*}

\subsection{Signal distribution}
\label{Sec:gwtc2.1_sig_kde}

The signal distributed is generated by fitting simulated software signals or \emph{injections} that are added to the detector noise, to a \ac{kde}. 
We use the same simulated signals as from the GWTC-3 analysis~\cite{ligo_scientific_collaboration_and_virgo_2021_5546676}, whose distribution is described in detail in Appendix~C.7 (the injection row of Table~X) of the GWTC-3 paper~\cite{LIGOScientific:2021djp}. 
In particular, the source masses are distributed as 
\begin{equation}
    \begin{split}
        p(m_1) & \propto m_1^{-2.35}, \\ 
        \quad p(m_2 \, | \, m_1) & \propto  m_2.
    \end{split} 
\end{equation}
The redshift distribution is flat in comoving volume, with a maximum redshift of $1.9$, assuming a $\Lambda$CDM cosmology~\cite{Planck:2015fie}. 
We restrict ourselves to injections with source-frame component masses $3 M_{\odot}$--$100 M_{\odot}$ to focus solely on \acp{bbh}.  
The simulated population is not a close match to the inferred astrophysical population~\cite{LIGOScientific:2021psn,Callister:2023tgi}. 
However, it shares broad characteristics and is simple to use. 
Therefore it suffices for our illustrative calculation. 
Using a population model that more closely resembles the underlying population should enable more accurate calculation of the true alarm rate, and hence \pastro{}.

\boldtext{In any KDE-based analysis, the choice of the KDE bandwidth is important, especially for multidimensional distributions with intricate shapes. 
In our illustrative analysis, the bandwidth was picked manually by checking that the \ac{kde} approximates well the distribution of injections. 
We verified that the one-dimensional cumulative distributions of the \ac{kde} and the injection set match, and also cross-checked the \ac{kde} against a \ac{kde} using the Silverman rule of thumb~\cite{silverman1986density}.
Ultimately, a realistic application would need a more sophisticated fitting scheme, for example, \acp{kde} fit using an iterative approach~\cite{Sadiq:2023zee} or a machine-learning approach such as using normalizing flows.}  

The top plot of Fig.~\ref{fig:joi_inj_kde} shows the joint distribution of signals that are found by both \pycbc{} and \gstlal, and Fig,~\ref{fig:marg_distribution} shows the distribution of signal triggers that are found only by one pipeline.

\subsection{Noise distribution}
\label{Sec:gwtc2.1_noise_kde}

The noise distribution is fit using data with a nonphysical time shift higher than the light travel time between detectors to remove correlations between any real signals. 
Generating this data is computationally expensive, therefore, in this proof-of-concept analysis, we use O3a data with a single time shift where LIGO Livingston and Virgo data streams were shifted by $0.62831$ and $0.31415~\mathrm{s}$, respectively, in relation to LIGO Hanford.

To define corresponding noise triggers between the pipelines, we match the times of the time-shifted triggers within a window of $1~\mathrm{s}$, comparable to the typical window used for grouping triggers in O3~\cite{LIGOScientific:2021djp}. 
This procedure gives us $256$ \gstlal{} triggers and $348$ \pycbc{} triggers that pass the \ac{far} threshold of $2~\mathrm{day}^{-1}$; of these, only $12$ triggers are common between pipelines. 
For a unified \pastro{} analysis intended to be used in production of a \ac{gw} catalog, a larger set of noise triggers \boldtext{would be required}.

The small number of noise triggers common between the pipelines makes it difficult to accurately reconstruct the distribution. 
However, as our primary goal is to illustrate unified \pastro{} method, rather than to produce a full set of results, these triggers are sufficient for completing a demonstrative calculation. 
To make use of the small number of triggers that are common between the pipelines, we adopt a bootstrapping procedure to generate more triggers. 
To do this, we first fit triggers that pass a $10~\mathrm{day}^{-1}$ cut to a \ac{kde}, and draw $10^4$ triggers that pass a $2~\mathrm{day}^{-1}$ cut. 
This corresponds to doing $\approx 17$ time slides in total. 
To avoid overfitting the noise triggers, we choose to employ a truncated Gaussian fit to them, instead of a KDE.
The assumption of a Gaussian is purely for simplicity, and a more careful choice of estimating the noise distribution would be needed for a proper analysis. 
The single-pipeline noise triggers are fit using a truncated Gaussian with mean $\mu = \beta(2~\mathrm{day}^{-1})$ and a standard deviation calculated from the noise triggers. The joint noise distribution is fit with a $2$-dimensional truncated Gaussian with $\mu = \beta(2~\mathrm{day}^{-1})$ and the covariance matrix calculated from the noise triggers. 
Figure~\ref{fig:joi_noi_kde} shows the distribution of joint triggers, while Fig.~\ref{fig:marg_distribution} shows the single-pipeline triggers in this bootstrapped data set.

The sparseness of noise triggers means that the noise distribution is not accurately reconstructed. 
This is liable to give biased values for \pastro{}: where the noise likelihood is underestimated, \pastro{} will be underestimated, and where the noise likelihood is overestimated, \pastro{} will be underestimated. 
However, the goal of this application is to show that the unified \pastro{} formalism yields sensible results to illustrate what would be needed for future analyses. 
A realistic application of this process to GW data will likely require multiple time shifts or some other process to generate a larger number of high-fidelity noise triggers, and then a careful reconstruction of the shape of the noise distribution.

\subsection{Computing a unified \pastro}

\begin{table*}
    \centering
    \begin{tabular*}{0.95\textwidth}{c  @{\extracolsep{\fill}} c c c c c c } 
  \hline 
 \hline 
&& \multicolumn{2}{c}{\textsc{GstLAL}} & \multicolumn{2}{c}{\textsc{PyCBC}} \\
\cline{3-4} \cline{5-6}GPS time & GW name &  FAR $(\text{yr}^{-1})$ & $\beta$ & FAR $(\text{yr}^{-1})$ & $\beta$ & Unified $p_{\rm astro}$ \\  
 \hline 
$1238782700.0$ &GW$190408\_181802$ & $2.1\times 10^{-15}$& $38.4$ & $2.5\times 10^{-04}$& $12.9$ & $> 0.99$ \\ 
$1239082262.0$ &GW$190412\_053044$ & $1.9\times 10^{-27}$& $66.1$ & $1.1\times 10^{-04}$& $13.7$ & $> 0.99$ \\ 
$1240327333.0$ &\-- & $9.1\times 10^{-01}$& $4.7$ & $4.2\times 10^{+01}$& $1.2$ & 0.98 \\ 
$1240944862.0$ &GW$190503\_185404$ & $2.3\times 10^{-06}$& $17.6$ & $3.8\times 10^{-02}$& $7.9$ & $> 0.99$ \\ 
$1241108686.0$ &\-- & $1.7\times 10^{+01}$& $1.9$ & - & - & 0.85 \\ 
$1241719652.0$ &GW$190512\_180714$ & $7.7\times 10^{-12}$& $30.2$ & $1.1\times 10^{-04}$& $13.7$ & $> 0.99$ \\ 
$1241816086.0$ &GW$190513\_205428$ & $1.3\times 10^{-05}$& $15.8$ & - & - & $> 0.99$ \\ 
$1242107479.0$ &GW$190517\_055101$ & $4.5\times 10^{-03}$& $10.0$ & $9.4\times 10^{-03}$& $9.3$ & $> 0.99$ \\ 
$1242315362.0$ &GW$190519\_153544$ & $2.2\times 10^{-06}$& $17.6$ & $1.0\times 10^{-04}$& $13.8$ & $> 0.99$ \\ 
$1242442967.0$ &GW$190521\_030229$ & $2.0\times 10^{-01}$& $6.2$ & $4.4\times 10^{-01}$& $5.4$ & $> 0.99$ \\ 
$1242459857.0$ &GW$190521\_074359$ & $5.0\times 10^{-33}$& $79.0$ & $1.8\times 10^{-05}$& $15.5$ & $> 0.99$ \\ 
$1242984073.0$ &GW$190527\_092055$ & $2.3\times 10^{-01}$& $6.1$ & - & - & $> 0.99$ \\ 
$1243533585.0$ &GW$190602\_175927$ & $1.1\times 10^{-07}$& $20.6$ & $2.9\times 10^{-01}$& $5.9$ & $> 0.99$ \\ 
$1243926576.0$ &\-- & $1.2\times 10^{+01}$& $2.2$ & $6.4\times 10^{+02}$& $0.1$ & 0.72 \\ 
$1243985856.0$ &\-- & $3.2\times 10^{+02}$& $0.3$ & $2.6\times 10^{+02}$& $0.3$ & 0.59 \\ 
$1245221073.0$ &\-- & $1.2\times 10^{+02}$& $0.6$ & $1.8\times 10^{+02}$& $0.4$ & 0.53 \\ 
$1245874666.0$ &\-- & $6.2\times 10^{+02}$& $0.1$ & $2.2\times 10^{+02}$& $0.4$ & 0.80 \\ 
$1246048404.0$ &GW$190701\_203306$ & $5.7\times 10^{-03}$& $9.8$ & $6.4\times 10^{-02}$& $7.4$ & $> 0.99$ \\ 
$1246385767.0$ &\-- & $5.3\times 10^{+00}$& $3.0$ & - & - & $> 0.99$ \\ 
$1246487219.0$ &GW$190706\_222641$ & $5.0\times 10^{-05}$& $14.5$ & $3.7\times 10^{-04}$& $12.5$ & $> 0.99$ \\ 
$1246779793.0$ &\-- & $1.0\times 10^{+01}$& $2.4$ & - & - & 0.99 \\ 
$1246849694.0$ &\-- & $2.7\times 10^{+00}$& $3.6$ & - & - & $> 0.99$ \\ 
$1247616534.0$ &GW$190720\_000836$ & $4.4\times 10^{-08}$& $21.5$ & $1.4\times 10^{-04}$& $13.5$ & $> 0.99$ \\ 
$1248242631.0$ &GW$190727\_060333$ & $2.7\times 10^{-10}$& $26.6$ & $5.6\times 10^{-03}$& $9.8$ & $> 0.99$ \\ 
$1248331528.0$ &GW$190728\_064510$ & $5.4\times 10^{-16}$& $39.8$ & $8.2\times 10^{-05}$& $14.0$ & $> 0.99$ \\ 
$1248617394.0$ &GW$190731\_140936$ & $3.3\times 10^{-01}$& $5.7$ & - & - & $> 0.99$ \\ 
$1249479778.0$ &\-- & $6.6\times 10^{+00}$& $2.8$ & - & - & $> 0.99$ \\ 
$1250620378.0$ &\-- & $5.2\times 10^{+00}$& $3.0$ & - & - & $> 0.99$ \\ 
$1251009263.0$ &GW$190828\_063405$ & $5.0\times 10^{-27}$& $65.2$ & $8.5\times 10^{-05}$& $14.0$ & $> 0.99$ \\ 
$1251010527.0$ &GW$190828\_065509$ & $3.5\times 10^{-05}$& $14.9$ & $2.8\times 10^{-04}$& $12.8$ & $> 0.99$ \\ 
$1251588283.0$ &\-- & $1.6\times 10^{+01}$& $2.0$ & - & - & 0.92 \\ 
$1251926900.0$ &\-- & $1.0\times 10^{+01}$& $2.4$ & - & - & $> 0.99$ \\ 
$1252627040.0$ &GW$190915\_235702$ & $7.8\times 10^{-06}$& $16.4$ & $6.8\times 10^{-04}$& $11.9$ & $> 0.99$ \\ 
$1252699636.0$ &GW$190916\_200658$ & $1.2\times 10^{+01}$& $2.2$ & - & - & 0.99 \\ 
$1252756008.0$ &GW$190917\_114630$ & $6.6\times 10^{-01}$& $5.0$ & - & - & $> 0.99$ \\ 
$1252939489.0$ &\-- & $9.0\times 10^{+00}$& $2.5$ & - & - & $> 0.99$ \\ 
$1252987339.0$ &\-- & $5.1\times 10^{+01}$& $1.1$ & $8.0\times 10^{+01}$& $0.8$ & 0.65 \\ 
$1253326744.0$ &GW$190924\_021846$ & $5.0\times 10^{-10}$& $26.0$ & $8.2\times 10^{-05}$& $14.0$ & $> 0.99$ \\ 
$1253509434.0$ &GW$190926\_050336$ & $1.1\times 10^{+00}$& $4.5$ & - & - & $> 0.99$ \\ 
$1253755327.0$ &GW$190929\_012149$ & $1.5\times 10^{-01}$& $6.5$ & $1.2\times 10^{+02}$& $0.6$ & $> 0.99$ \\ 
$1253885759.0$ &GW$190930\_133541$ & $4.3\times 10^{-01}$& $5.5$ & $1.8\times 10^{-02}$& $8.6$ & $> 0.99$ \\ 
 \hline 
 \hline 
\end{tabular*}
    \caption{Triggers with a unified $\pastro \geq 0.5$ from our illustrative analysis.   
    The triggers that have $\pastro{} \geq 0.5$ in at least one pipeline in GWTC-2.1~\cite{LIGOScientific:2021usb} are shown in the second column. Also listed are the \acp{far} of the triggers from the \gstlal{} and \pycbc{} pipelines. 
    \boldtext{The GW name of the triggers also recovered with \pastro{}$\geq 0.5$ by at least one pipeline in GWTC-2.1 is given in the second column.} 
    These results illustrate the properties of the unified \pastro{} method, but a larger number of noise triggers, and more accurate population models, would be needed to obtain reliable quantitative results. }
    
    \label{Tab:sig_pastro}
\end{table*}

After restricting ourselves to ones that are at least Hanford-Livingston coincident, and have been found by at least one of \gstlal{} or \pycbc, we are left with $553$ on-source triggers from the GWTC-2.1 O3a results~\cite{LIGOScientific:2021djp}. 
Using the signal and noise models described above, we can now proceed to calculate a unified \pastro{} for these triggers. 

First, through Eq.~\eqref{Eq:FGMC_estimate}, we estimate $\lambs{}$ and $\lambn{}$ for the joint analysis. 
We do this using the Markov-chain Monte Carlo sampler \textsc{emcee}~\cite{emcee:2013}. 
We use uniform priors on the count parameters, $\lambs{} \in [0, 1000]$ and  $\lambn{} \in [0, 1000]$. 
The triggers are then scored against the appropriate distribution to calculate $ p({x_i} | \mathcal{S} )$ and $ p({x_i} | \varnothing)$.
 
Figure~\ref{fig:fgmc} shows the posterior distribution of $\lambs{}$ and $ \lambn{}$, with a median $\lambs{}$ of $53$.
Table~\ref{Tab:sig_pastro} lists the $41$ triggers that have unified $\pastro\geq 0.5$. 
Most triggers are reported by both \gstlal{} and \pycbc{}, but we do have a few triggers that are detected by \gstlal{} alone, which corresponds to the high $\beta$ tail of the on-source triggers in the top panel of Fig.~\ref{fig:marg_distribution}. 
This is qualitatively consistent with the $44$ triggers found with a $\pastro\geq 0.5$ by at least one pipeline in GWTC-2.1 analysis~\cite{LIGOScientific:2021djp}.

In Table~\ref{Tab:sig_pastro}, we recover $26$ triggers that are also reported with $\pastro{}>0.5$ in at least one search pipeline in the GWTC-2.1 analysis~\cite{LIGOScientific:2021djp} (the corresponding GW identification of these triggers has been given for easy identification) while the remaining $15$ triggers were below this threshold. 
Inspecting the $\beta$ values of these triggers show that they are usually low, and comparing with the noise distributions, it is plausible that they could be consistent with noise if some of the modeling assumptions are relaxed. 
The simplified modeling of the noise distribution is expected to lead to the promotion of some noise triggers while suppressing some real signals.

\begin{table*}
    \centering
    \begin{tabular*}{0.95\textwidth}{c  @{\extracolsep{\fill}} c c c c c c } 
  \hline 
 \hline 
&& \multicolumn{2}{c}{\textsc{GstLAL}} & \multicolumn{2}{c}{\textsc{PyCBC}} \\
\cline{3-4} \cline{5-6}GPS time & GW name &  FAR $(\text{yr}^{-1})$ & $\beta$ & FAR $(\text{yr}^{-1})$ & $\beta$ & Unified $p_{\rm astro}$ \\  
 \hline 
$1239917954.0$ &GW$190421\_213856$ & $2.8\times 10^{-03}$& $10.5$ & $5.9\times 10^{+00}$& $2.9$ & $< 0.01$ \\ 
$1246527224.0$ &GW$190707\_093326$ & $2.7\times 10^{-15}$& $38.2$ & $9.7\times 10^{-06}$& $16.1$ & $< 0.01$ \\ 
$1248834439.0$ &GW$190803\_022701$ & $7.3\times 10^{-02}$& $7.2$ & $8.1\times 10^{+01}$& $0.8$ & $< 0.01$ \\ 
 \hline 
 \hline 
\end{tabular*}
    \caption{\boldtext{Triggers with a unified $\pastro < 0.5$ in our illustrative analysis, but with a $\pastro \geq 0.5$ in either \gstlal{} or \pycbc{} in GWTC-2.1~\cite{LIGOScientific:2021usb}}}
    
    \label{Tab:demoted_pastro}
\end{table*}

In the case of joint triggers there is a heavy weight in favor of signal triggers as discussed in Sec.~\ref{sec:realdata}. 
The exact quantitative results might be susceptible to small-number statistics, and it is plausible that some of these would be down weighted if we had a more complete noise distribution from more time slides. 
However, the results show the expected behavior that having multiple pipelines find a candidate when their noise distributions are largely uncorrelated, increases our certainty that a candidate is real.

\boldtext{In Table~\ref{Tab:demoted_pastro}, we show the list of triggers with a unified $\pastro{}<0.5$, but with a $\pastro{} \geq 0.5$ in either \gstlal{} or \pycbc{} in GWTC-2.1~\cite{LIGOScientific:2021usb}. 
While the correlation between pipelines increases \pastro{} for some candidates, the unified analysis also down-weights certain triggers, in particular those with highly asymmetric $\beta$ values. 
This is because the joint simulated-signal trigger distribution has a strong correlation between the $\beta$ (and thereby the \ac{far}) values of two pipelines; this is expected since both are matched filter pipelines using \ac{cbc} templates.}
\boldtext{For example, GW190803\_022701 has \gstlal{} and \pycbc{} \acp{far} of $7.3 \times 10^{-2}~\mathrm{yr}^{-1}$ and $81~\mathrm{yr}^{-1}$, respectively. 
Therefore, while the \gstlal{} significance of the trigger is high, the asymmetric \acp{far} in the two pipelines gives it a lower weight in the unified analysis. 
In our case, the low \pastro{} can potentially be explained by the choice of a Gaussian for the noise distribution with heavy tails at large $\beta$. 
The uncertainties involved in the noise and the signal distributions in this illustrative analysis, and the fact that we do not include other search pipelines, mean that we should not rule out the triggers in Table~\ref{Tab:demoted_pastro}. Nevertheless, the results demonstrate that there is information that is uniquely captured by a joint analysis of multiple pipelines. 
In practice, we would expect that any high-significance candidate that is a clear outlier in the signal distribution would warrant further investigation to understand why the pipelines differ in their response.}

Finally, many of the triggers in Table~\ref{Tab:sig_pastro} have a \pastro{} close to $1$. 
This is likely an overestimate, due to the sparseness of the noise distribution which can lead to an artificially low noise likelihood in Eq.~\eqref{Eq:joint_pastro_marg}. 
We equally expect that some of the low-probability candidates, not shown in the table, have underestimated \pastro{} for the same reason. 
The specific values we get are also dependent on the strong modeling assumptions made for the noise distribution. 
A more realistic analysis would probably require adequate non-parametric modeling of the noise distribution such that any structure in the parameter space can be identified. 
While it is worth bearing these limitations in mind, both Table~\ref{Tab:sig_pastro} and Fig.~\ref{fig:fgmc} qualitatively demonstrate consistency of the results with GWTC-2.1, indicating that even with simpler modeling choices this method can yield sensible overall results.

\section{Discussion and conclusion}
\label{Sec:Discussion}

We have developed a statistical formalism for combining information from multiple search pipelines to calculate a unified \pastro{}. 
We first demonstrated this formalism using a simple toy model, showing that it can consistently combine information and even account for biases in search pipelines. 
We then applied the framework to O3a data, using triggers from \gstlal{} and \pycbc{}, demonstrating how correlations between pipelines may update our understanding of candidates, but highlighting the importance of using accurate models for the signal and noise populations. 

Currently, multiple search pipelines are used to identify interesting candidates, each with its own strengths. 
A unified \pastro{} can potentially reduce confusion in the interpretation of triggers found by different pipelines and can help in using marginal triggers for subsequent \ac{gw} analyses. 
We have shown that certain types of correlations between pipelines can inform the significance that the unified analysis assigns to a trigger. 
Similarly, combining the search pipelines' results mitigates the need to calculate an effective trials factor to correct the individual-pipeline \acp{far} to account for repeated analysis of the same data. 
Calculating this trials factor is generally nontrivial. 
Our formalism can naturally and consistently account for these considerations since it calculates how triggers in one pipeline are correlated with those in another pipeline. 

While we have demonstrated that this method can estimate signal and noise counts, and a unified \pastro{} in a sensible and consistent manner, the paucity of noise triggers is a clear computational bottleneck. 
The scarcity of these triggers can be understood by considering the \ac{far} thresholds used. The $2~\mathrm{day}^{-1}$ threshold used means that we will be limited to a few hundred noise triggers in each pipeline per a six month run. 
Therefore, depending on the structure in the joint noise correlations, we would probably need $\mathcal{O}(10$--$100)$ time slides to obtain an accurate representation of the noise distribution. 
The number of noise triggers that are common between pipelines will be only a fraction of the total number of noise triggers, and if the noise backgrounds are distinct, then there will only be a small fraction in common. 
This may indicate a need for multiple coordinated time shifts between pipelines to build up the noise distribution. One can also consider other methods of building noise distributions such as modeling the joint noise distribution and drawing from it, similar in philosophy to the \gstlal{} pipeline~\cite{Cannon:2012zt}.

However, while the low number of common noise triggers \boldtext{does pose} a difficulty for reconstructing the shape of the noise distribution, it also highlights the importance of considering the number of candidates in common between pipelines: if common noise triggers are rare, but common signal triggers are frequent, then a candidate being found by multiple pipelines should increase its \pastro{}. 
We already see this in the analysis done here where a joint trigger is about $100$ times more likely under the signal hypothesis. 

If suitable data products (injection sets and noise triggers) are available, this formalism can be readily applied to include other search pipelines, notably those used by the LVK such as \textsc{MBTA}~\cite{Adams:2015ulm, Aubin:2020goo} and \textsc{cWB}~\cite{klimenko:2004, Klimenko:2011hz, Klimenko:2015ypf}, as well as pipelines developed for external analysis of public data~\cite{Venumadhav:2019tad}. 
This would enable the construction of a single \ac{gw} catalog accounting for all analyses. 
\boldtext{A strength of this formalism when constructing joint catalogs is that it would differentially weigh search pipelines, depending on their precision and accuracy by taking into account the shape of the joint simulated-signal distribution. Hence, as long as the simulated-signal distribution sufficiently reflects the true distribution of GW sources, it should be possible to correct for pipelines missing signals over some region of parameter space, or producing spurious triggers in another}.

The addition of \textsc{cWB} (and other minimally modeled pipelines~\cite{Lynch:2015yin,Kanner:2015xua}) \boldtext{to catalog production} could be extremely useful, as it can be sensitive in regions of the \ac{cbc} parameter space where modeled searches do not perform well, such as eccentric binaries~\cite{LIGOScientific:2019dag}, in addition to non-\ac{cbc} transient sources. 
The sensitivity to non-\ac{cbc} sources is a complication for current analyses, as the \pastro{} calculations are done assuming only \ac{cbc} sources (discussed in Appendix~F of the GWTC-3 paper~\cite{LIGOScientific:2020ibl}). 
Non-\ac{cbc} source have a lower true alarm rate than \ac{cbc} sources, and hence a lower \pastro{} at a given \ac{far}; consequently, misidentifying a trigger as \ac{cbc} will lead to an overestimate of its \pastro{}.
To mitigate this, LVK analyses have imposed an additional criterion for \textsc{cWB} candidates, that they must have a counterpart trigger from a template-based \ac{cbc} pipeline~\cite{LIGOScientific:2018mvr,LIGOScientific:2021djp}. 
This is effectively an approximation to the unified \pastro{} framework, acknowledging that for real \ac{cbc} signals there would probably be correlation between pipeline results. 
This could be put on a more rigorous basis by explicitly using out unified \pastro{} framework, considering the response of the pipelines to simulated \ac{cbc} and non-\ac{cbc} signals.

Addition of more pipelines can make the problem of noise triggers more acute, as the number of triggers that are common between three or more pipelines might be small and extrapolation difficult. 
However, it is probable that noise triggers in a multiple-pipeline space would be so rare that we can conservatively set the noise likelihood to a small limiting value for such a case.  

In this paper, we have only considered the simplest case of calculating a unified \pastro{}. 
In a future work, we will extend it to implement binning in the mass space to estimate the astrophysical probability that a system is a \ac{bbh}, a \ac{bns} or a \ac{nsbh}~\cite{Kapadia:2019uut}. 
Such an extension is in principle straightforward, where in addition to a statistic like the \ac{far} we also fit distribution of recovered template chirp masses for simulated \ac{bbh}, \ac{bns} and \ac{nsbh} signal. 
On-source triggers can then be scored against such two-dimensional distributions to give the corresponding likelihood. 
Searches are often optimized in different ways giving them differential sensitivity in specific parts of the mass space (e.g., \pycbc{} has an analysis tuned to \acp{bbh}~\cite{Nitz:2020oeq,LIGOScientific:2020ibl}). 
Calculating unified probabilities in conjugation with mass binning will allow us to fold in this differential sensitivity in a consistent, unbiased way in a way akin to the toy model example with \toypipe{3} and \toypipe{4} in Sec.~\ref{sec:toy-model}.

A key uncertainty in the calculation of \pastro{} is the form of the underlying source population. 
Errors in the assumed population translate to a misestimation of the true alarm rate, and hence \pastro{}. 
A way to mitigate this would be to infer the population simultaneously while calculating \pastro{}~\cite{Gaebel:2018poe,Galaudage:2019jdx, Roulet:2020wyq}. 
This additionally enables lower-significance candidates to be used in population inference. 
Using a unified \pastro{} makes it easier to assess the contamination fraction in a catalog based upon several pipelines, and hence would make it more convenient to perform such joint inference in the future. 

While we have only considered the application of our method to final search analyses performed offline, an extension to low-latency detections is also theoretically possible. 
By combining multiple pipelines we depend less on one pipeline and hence should be less susceptible to incorrect alerts and retractions. 
In combination with mass binning, this could be extremely useful for electromagnetic follow-up of \ac{gw} candidates~\cite{Lynch:2018yom}. 
Since many such follow-ups are often target-of-opportunity observations, improving the reliability of trigger information can be valuable in evaluating the proper usage of scarce telescope time. 
The relative scarcity of noise triggers could be a bigger computational issue in low latency, as the joint noise distributions will have to be continuously reevaluated at a reasonable cadence in order to account for the changing detector state~\cite{aLIGO:2020wna,Virgo:2022ypn}. 
Therefore, further work would be needed to identify methods that could potentially be used to reconstruct the noise distribution in low latency.  

Finally, Bayesian frameworks have been developed to assess the probability of multimessenger detections~\cite{Ashton:2017ykh,Bartos:2018jco,Veske:2020rxf,Piotrzkowski:2021hhy}. 
These calculate the probabilities associated with candidates being background noise or astrophysical signals from different sources or the same source. 
Our unified \pastro{} framework naturally feeds into these calculations. 
Furthermore, our framework would enable the extension of these calculations to consider how a nondetection in one messenger impacts the probability that a candidate in a counterpart messenger is real. 
Given the rich science that may result from a multimessenger discovery~\cite{LIGOScientific:2017ync,KAGRA:2013rdx,Margutti:2020xbo}, this may be a valuable avenue of future investigation.

The data used in this paper for the analysis of GWTC-2.1 triggers is available as a Zenodo repository~\cite{zenodo_pastro}.

\section*{Acknowledgements} \label{sec:acknowledgements}

We thank Will Farr, Vicky Kalogera and Surabhi Sachdev for useful discussions. 
We also thank Anarya Ray, Tom Dent \boldtext{and the anonymous referees} for helpful comments on the manuscript. 
S.B. was supported by National Science Foundation (NSF) Grant No. PHY-2207945. C.P.L.B. acknowledges support from Science and Technology Facilities Council (STFC) Grant No. ST/V005634/1. 
Z.D. acknowledges support from the CIERA Board of Visitors Research Professorship. 
L.T. is supported by the NSF through Grants No. OAC-2103662 and No. PHY-2011865. 
GSCD acknowledges the STFC for funding through Grants No. ST/T000333/1 and No. ST/V005715/1. 
The authors are grateful for computational resources provided by the LIGO Laboratory and supported by NSF Grants No. PHY-0757058 and No. PHY-0823459. This material is based upon work supported by NSF's LIGO Laboratory which is a major facility fully funded by the National Science Foundation. This research has made use of data obtained from the Gravitational Wave Open Science Center (\href{https://gwosc.org}{gwosc.org})~\cite{LIGOScientific:2023vdi}, a service of LIGO Laboratory, the LIGO Scientific Collaboration, the Virgo Collaboration, and KAGRA. LIGO Laboratory and Advanced LIGO~\cite{LIGOScientific:2014pky} are funded by the United States NSF as well as the STFC of the United Kingdom, the Max-Planck-Society (MPS), and the State of Niedersachsen/Germany for support of the construction of Advanced LIGO and construction and operation of the GEO\,600 detector~\cite{Dooley:2015fpa}. Additional support for Advanced LIGO was provided by the Australian Research Council. Virgo~\cite{VIRGO:2014yos} is funded, through the European Gravitational Observatory (EGO), by the French Centre National de Recherche Scientifique (CNRS), the Italian Istituto Nazionale di Fisica Nucleare (INFN) and the Dutch Nikhef, with contributions by institutions from Belgium, Germany, Greece, Hungary, Ireland, Japan, Monaco, Poland, Portugal, Spain. KAGRA~\cite{KAGRA:2020agh} is supported by Ministry of Education, Culture, Sports, Science and Technology (MEXT), Japan Society for the Promotion of Science (JSPS) in Japan; National Research Foundation (NRF) and Ministry of Science and ICT (MSIT) in Korea; Academia Sinica (AS) and National Science and Technology Council (NSTC) in Taiwan.
Corner plots were made with the Chainconsumer~\cite{Hinton:2016} package. 
This document carries the LIGO DCC number P2300107.

\bibliography{references}

\end{document}